\documentclass[lettersize,journal]{IEEEtran}
\usepackage{amsmath,amsfonts}
\usepackage{algorithmic}
\usepackage{algorithm}
\usepackage{array}
\usepackage[caption=false,font=normalsize,labelfont=sf,textfont=sf]{subfig}
\usepackage{textcomp}
\usepackage{stfloats}
\usepackage{url}
\usepackage{verbatim}
\usepackage{graphicx}
\usepackage{cite}
\usepackage[hidelinks,
linkcolor=black,
anchorcolor=black,
citecolor=black]{hyperref}
\begin{document}

\title{Affine Frequency Division Multiplexing for Communication and Channel Sounding: Requirements, Challenges, and Key Technologies}
\author{Yu Zhou, Chao Zou, Nanhao Zhou, Yanqun Tang, Xiaoying Zhang, Haoran Yin, Xiaoran Liu, Ruisi He,~\IEEEmembership{Senior Member,~IEEE}, Pan Tang,~\IEEEmembership{Member,~IEEE}, Weijie Yuan,~\IEEEmembership{Senior Member,~IEEE}, and Yong Zeng,~\IEEEmembership{Fellow,~IEEE}
\thanks{Yu Zhou, Chao Zou, Nanhao Zhou, Yanqun Tang, and Haoran Yin are with the School of Electronics and Communication Engineering, Sun Yat-sen University, China.}
\thanks{Xiaoying Zhang and Xiaoran Liu are with the College of Electronic Science and Technology,
National University of Defense Technology, China.}
\thanks{Ruisi He is with the School of Electronics and Information Engineering, Beijing Jiaotong University, China.}
\thanks{Pan Tang is with the State Key Lab of Networking and Switching Technology, Beijing University of Posts and Telecommunications, China.}
\thanks{Weijie Yuan is with the School of System Design and Intelligent Manufacturing, Southern University of Science and Technology, China.}
\thanks{Yong Zeng is with the National Mobile Communications Research Laboratory and the Frontiers Science Center for Mobile Information Communication and Security, Southeast University and also with the Purple Mountain Laboratories, China.}
}



\maketitle

\begin{abstract}
Channel models are crucial for theoretical analysis, performance evaluation, and deployment of wireless communication systems.
Traditional channel sounding systems are insufficient for handling the dynamic changes of channels in the next-generation space-air-ground-sea integrated networks (SAGSIN), which often results in outdated channel models that fail to provide reliable prior information for communication systems.
To address this challenge, this paper proposes an integrated channel sounding and communication (ICSC) method as a practical solution.
Unlike orthogonal frequency division multiplexing, affine frequency division multiplexing (AFDM) provides a full delay-Doppler representation of the channel, achieving optimal diversity in time-frequency doubly dispersive channels and effectively addressing the aforementioned challenges.
Thus, we investigate the fundamental principles of AFDM, showing how it enables simultaneous communication and channel sounding, and explore key performance metrics for both functionalities.
We also clarify the distinction and relationship between channel sounding, estimation, tracking and scatterer sensing.
Additionally, several potential application scenarios for AFDM-ICSC are explored.
Finally, we highlight the key challenges in implementing AFDM-ICSC, outline future research directions, and provide valuable insights for the continued development of this technology.
\end{abstract}

\section{Introduction}
\begin{figure*}[!t]
\centering
\includegraphics[width=1\textwidth]{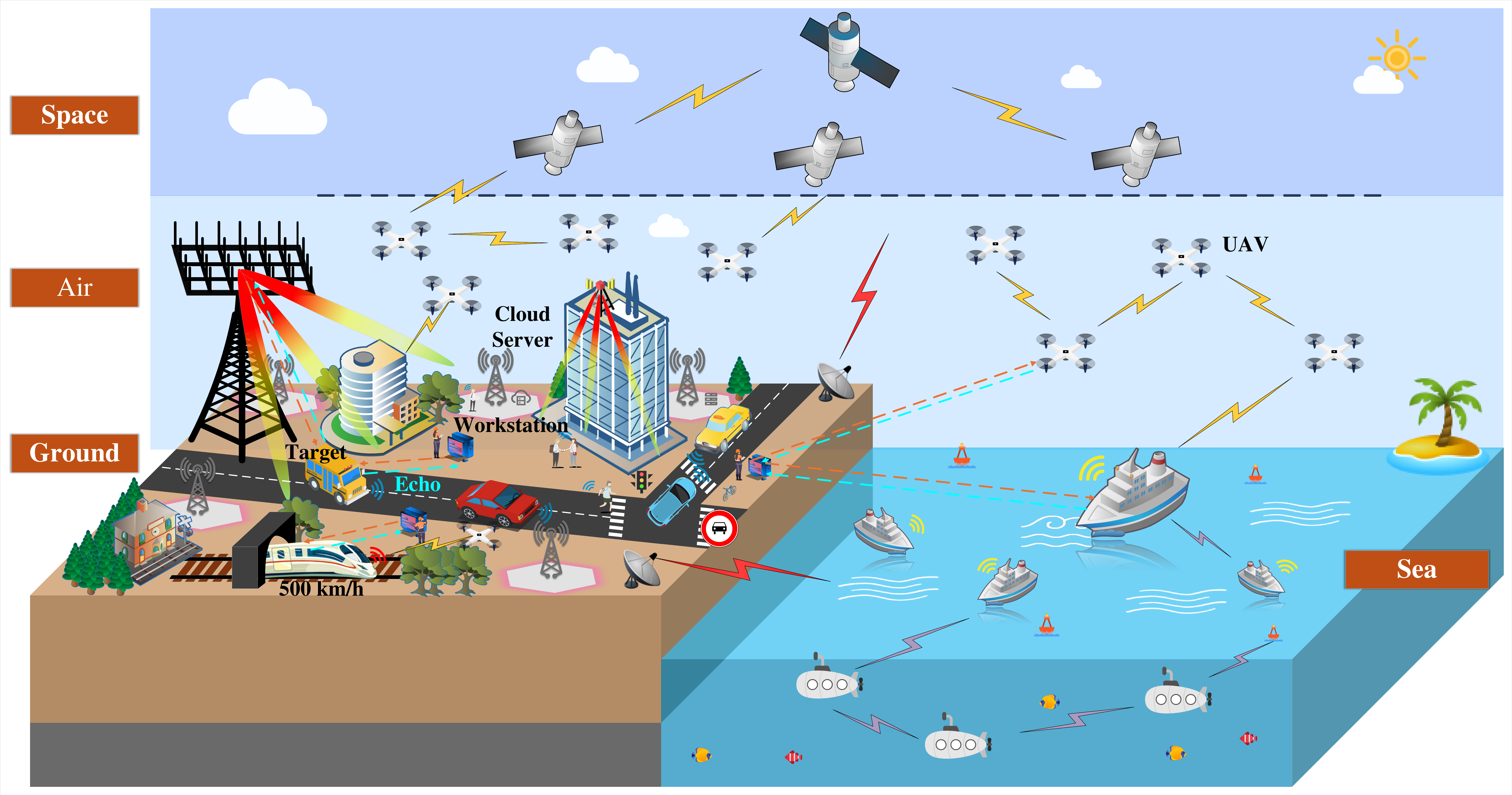}
\caption{An illustration of space-air-ground-sea integrated networks.}
\label{fig_1}
\end{figure*}

With the evolution of mobile communication systems, the study of the channel has progressively expanded across multiple dimensions, including scenarios, frequency, bandwidth, and spatial domains. 
Sixth-generation (6G) networks are envisioned to transition beyond traditional terrestrial communications, evolving into space-air-ground-sea integrated networks (SAGSIN) \cite{ref1}. 
These networks will incorporate satellites, unmanned aerial vehicles (UAVs), terrestrial ultra-dense networks, underground, maritime, and underwater communications, as illustrated in Fig. \ref{fig_1}.
The characteristics of channels fundamentally determine the performance limits of communication systems. 
Hence, a comprehensive understanding of channel characteristics in 6G networks is essential for guiding system design, enabling accurate performance evaluation, and optimizing algorithm development \cite{ref2}.

To achieve efficient and reliable communication in unknown scenarios, traditional methods typically involve using channel sounding systems based on sounding sequences like pseudo-noise (PN) or Zadoff-Chu (ZC) sequences to collect measurement data \cite{ref3}.
Subsequently, the data is analyzed to extract channel characteristics and construct a channel model, which is then used to optimize the transceiver design of the communication system.
However, this process imposes significant system overhead and fails to meet real-time communication requirements due to the complex and dynamic nature of the channel.
Furthermore, traditional channel sounding systems often rely on expensive and bulky high-precision equipment, making them cost-prohibitive and impractical for pre-deployment in many scenarios. 
This challenge is especially pronounced in satellite and underwater applications within SAGSIN, where spatial constraints, extreme environmental conditions (e.g., vacuum and high radiation for satellites; high pressure and restricted access for underwater), and miniaturization requirements render the deployment of such equipment highly unfeasible.
Consequently, traditional channel sounding systems are difficult to adapt to the complex and diverse application scenarios of 6G networks. 
In this regard, integrated channel sounding and communication (ICSC) offers an efficient and energy-saving solution \cite{refICSC}.
This approach enables real-time channel estimation using reference signals and conducts channel sounding through statistical analysis, thereby eliminating the need for expensive dedicated sounders.
Leveraging widespread communication infrastructure, ICSC reduces deployment costs and supports large-scale, long-term channel data collection.
Its adaptive sounding mechanism integrates seamlessly with communication systems, providing a scalable and flexible solution for complex and heterogeneous environments.

Orthogonal frequency division multiplexing (OFDM), widely employed in current fourth-generation (4G) and fifth-generation (5G) networks, offers excellent communication performance in time-invariant frequency-selective channels.
Training symbols and reference signals in the physical layer frames of Long-Term Evolution (LTE) and 5G New Radio (5G-NR) protocols are utilized to extract channel state information (CSI) \cite{ref4,ref5}.
This extraction facilitates the estimation of multipath parameters, allowing detailed characterization of channel properties.
However, 6G networks are required to support higher data rates and increasingly complex scenarios.
Moreover, 6G channels exhibit pronounced non-stationarity across space, time, and frequency domains due to larger antenna arrays, higher mobility, and wider bandwidths \cite{ref2}, resulting in space-, time-, and frequency-selective fading, respectively.
Notably, severe Doppler spread caused by high mobility compromises the orthogonality of OFDM subcarriers, significantly degrading communication performance.
Consequently, OFDM cannot meet the stringent demands of 6G's complex and dynamic scenarios, emphasizing the urgent need for innovative waveform designs and transformative technologies.

As a result, several new waveforms have emerged in recent years.
One of the most popular methods is orthogonal time frequency space (OTFS) modulation \cite{ref7}, which leverages the inverse symplectic finite Fourier transform (ISFFT) to modulate a two-dimensional grid of information symbols directly in the delay-Doppler domain, effectively capturing the quasi-static nature of the channel.
Besides, by exploiting the super-resolution of large antenna arrays, delay-Doppler alignment modulation (DDAM) has been proposed \cite{ref8}.
Another alternative is orthogonal chirp division multiplexing (OCDM), a chirp-based waveform with a fixed chirp rate determined by the discrete Fresnel transform (DFnT) \cite{ref9}.
However, the fixed chirp rate limits the ability to achieve full channel diversity.

Affine frequency division multiplexing (AFDM) is a novel multicarrier waveform that maps information symbols onto a set of orthogonal chirps using the discrete affine Fourier transform (DAFT), which generalizes the DFnT used in OCDM and the discrete Fourier transform (DFT) used in OFDM \cite{ref10,ref11}.
Unlike OCDM, which uses a fixed chirp rate defined by the DFnT, AFDM introduces two tunable parameters that allow flexible adaptation to doubly selective channels in both time and frequency domains, thereby enhancing its applicability across diverse scenarios.
This parameterized structure not only offers excellent backward compatibility for AFDM, ensuring seamless integration with OFDM, but also enhances its adaptability to complex channel conditions, resulting in significantly better communication performance, particularly in mitigating sensitivity to carrier frequency offset compared to OFDM.
Moreover, the two tunable parameters endow AFDM with inherent security capabilities, providing enhanced flexibility for physical layer security \cite{ref12,ref121}.
In contrast to the two-dimensional modulation in OTFS, AFDM employs a one-dimensional modulation structure, reducing the implementation complexity.
Furthermore, it offers clear advantages in channel estimation and multi-user access overhead, thereby achieving higher system throughput \cite{ref13}.
With these outstanding features, AFDM emerges as a promising and reliable candidate waveform for next-generation wireless communication systems.
To meet the dual demands of reliable communication and precise channel sounding in future 6G environments, integrating AFDM with ICSC is essential.

This article provides a comprehensive overview of AFDM-ICSC.
To facilitate a clearer understanding of ICSC, we first clarify the distinctions and interrelationships among channel sounding, estimation, tracking, and scatterer sensing.
Then, we present the architecture of AFDM-ICSC and outline the principle of parametric design.
Subsequently, potential applications and opportunities are outlined. 
Finally, we explore future research directions and key challenges.

\section{Channel sounding, estimation, tracking, and scatterer sensing}
\begin{figure*}[!t]
\centering
\includegraphics[width=0.98\textwidth]{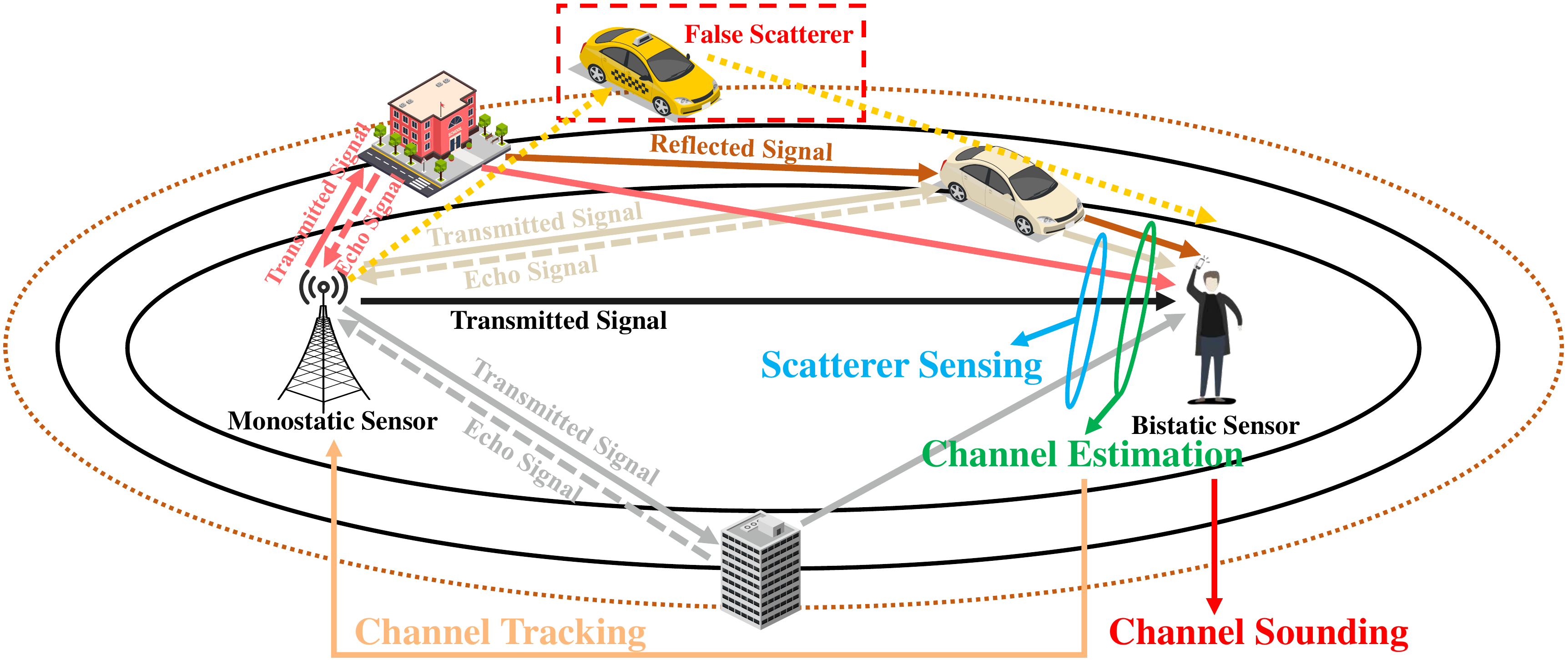}
\caption{An illustration of the communication and sounding scenarios involved in an ICSC system.}
\label{fig_2}
\end{figure*}

Characterizing the propagation environment via channel sounding is a key premise for stable and reliable wireless communication.
Channel sounding aims to extract the statistical characteristics of the channel, encompassing large-scale fading (e.g., path loss and shadowing) and small-scale fading such as the power delay profile (PDP), Doppler power spectrum (DPS), root mean square (RMS) delay spread, and RMS Doppler spread. 
A channel model based on these characteristics can be developed to support system design, performance evaluation, and the provision of prior channel information to the receiver.

As channel models are inherently statistical and prior, practical communication systems rely on reference signals to estimate the channel in real time. 
Specifically, the channel impulse response (CIR) characterizes the time-domain behavior of the channel, while the channel frequency response (CFR), as its Fourier transform, captures the frequency-domain characteristics. 
In most systems, CSI typically refers to the subcarrier-sampled CFR, containing amplitude and phase, and is essential for equalization and demodulation.
Depending on whether the channel is quasi-static or time-varying, the estimation results are not only used for local demodulation but may also be fed back to the transmitter (e.g., base station) for channel tracking and adaptive resource allocation.
For example, in quasi-static conditions, reference signal overhead can be minimized, while rapidly time-varying channels require denser reference signals to ensure estimation accuracy.
Moreover, assuming channel reciprocity, the base station can track the downlink channel based on the uplink channel estimation results.
In ICSC, instantaneous channel estimates can be used for signal equalization at the receiver, fed back to the transmitter for real-time tracking, or statistically processed over time to extract channel characteristics for sounding.

Scatterer sensing aims to estimate the distance, velocity, and angle of scatterers using wireless signals affected by the propagation environment.
As illustrated in Fig. \ref{fig_2}, in a bistatic scenario, the received signal comprises five components: a direct path from the transmitter, three single-reflection paths from scatterers, and one double-reflection path (the brown path, where the signal reflects off the pink house and then the white car).
The sensing objective is to estimate the parameters of the transmitter and the three true scatterers.
Specifically, four valid signal paths (highlighted with blue circles) must be identified and extracted for further processing.
However, the double-reflection path corresponds to a false scatterer (yellow car), whose total propagation distance coincides with that of a hypothetical single-reflection path (yellow dashed line), resulting in a false scatterer at the receiver that may interfere with accurate sensing.
In a monostatic scenario, the sensor estimates the scatterer parameters by analyzing the echoes reflected from its own transmitted signals.

In particular, integrated sensing and communication (ISAC) focuses on sensing specific scatterers or targets of interest within the channel. 
In contrast, ICSC aims to comprehensively characterize the multipath structure and temporal variations of the channel, requiring detailed modeling of all scatterers that impact the signal.
Moreover, instantaneous channel estimates can be fed back to the transmitter for channel tracking, while their statistical analysis over time enables channel sounding for propagation environment characterization.
The relationship among channel sounding, estimation, tracking, and scatterer sensing is illustrated in Fig. \ref{fig_2}.

\section{Fundamentals of AFDM-ICSC}
We first present the architecture of AFDM-ICSC, followed by a comprehensive analysis of its parameter design principles.

\subsection{AFDM-ICSC}
\begin{figure*}[!t]
\centering
\includegraphics[width=0.97\textwidth]{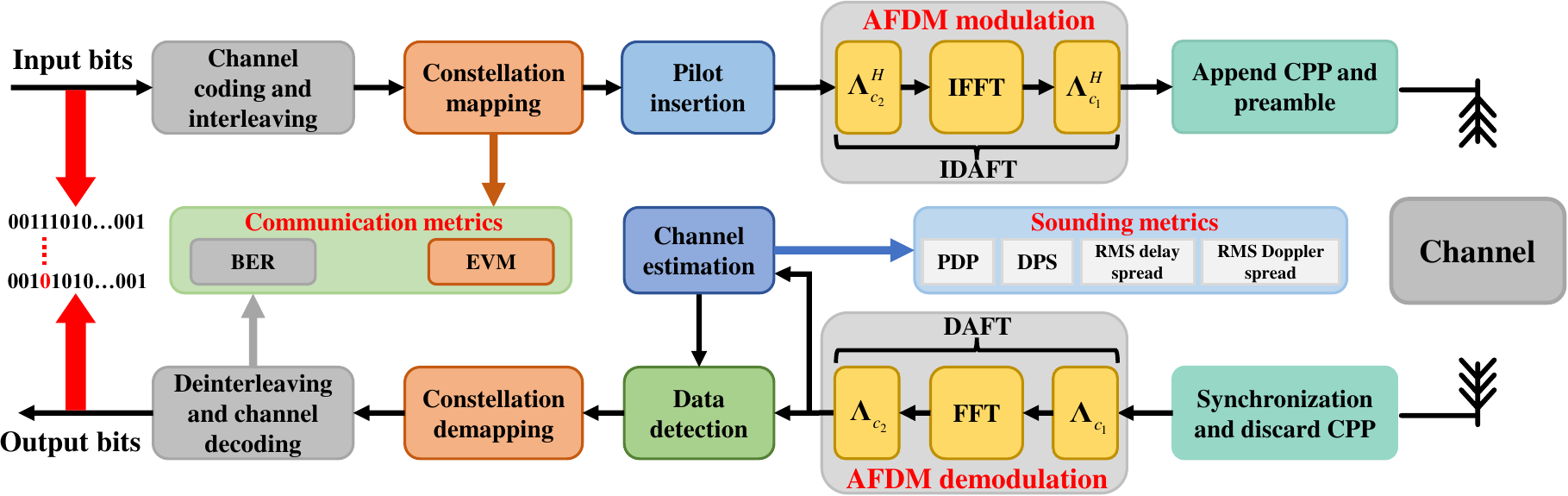}
\caption{The architecture of AFDM-ICSC.}
\label{fig_3}
\end{figure*}
\begin{figure}[!t]
\centering
\subfloat[\textrm{AFDM}]{
\includegraphics[width=0.45\textwidth]{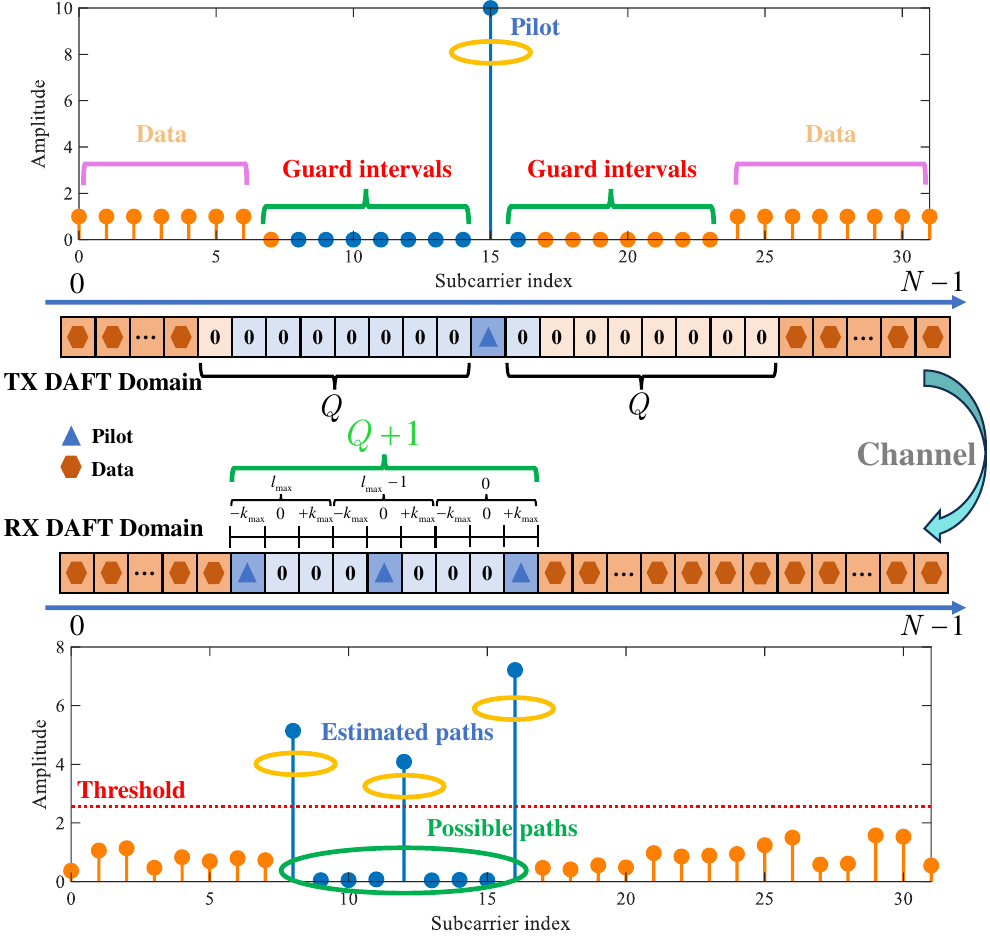}
\label{fig_4a}}\\
\subfloat[\textrm{OTFS}]{
\includegraphics[width=0.45\textwidth]{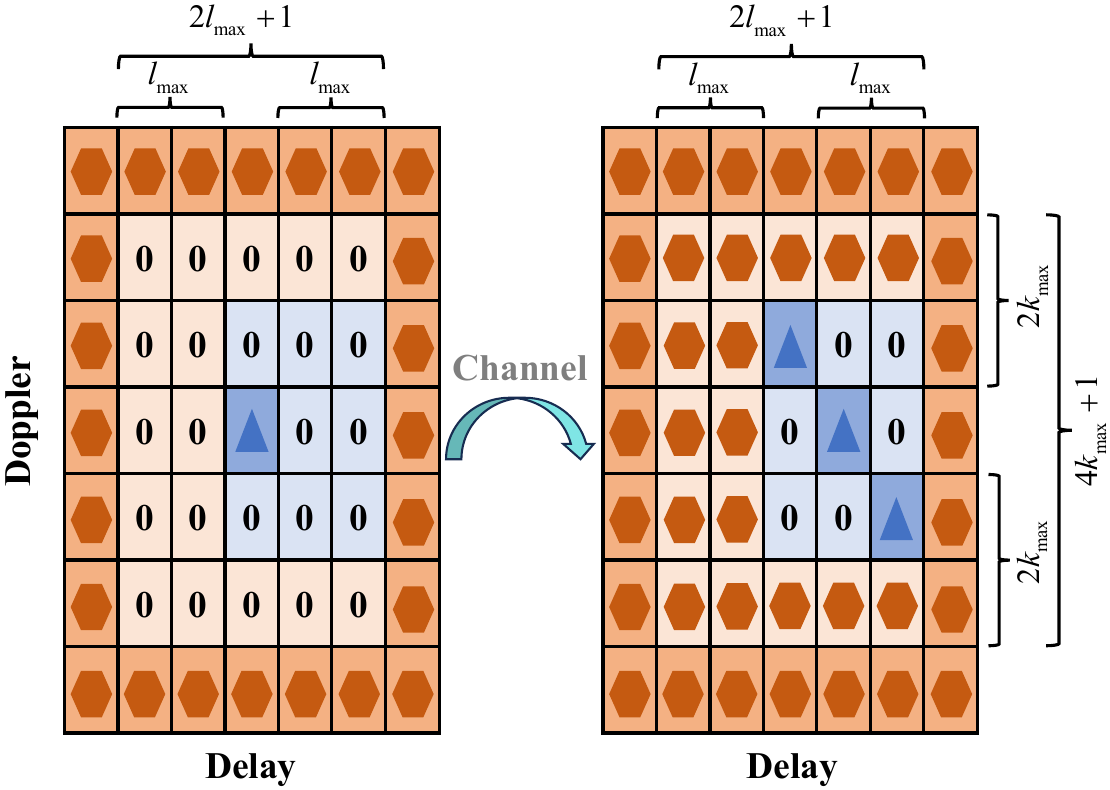}
\label{fig_4b}}\\
\caption{A schematic diagram of embedded pilot channel estimation ($N=32$, $P=3$, $l_{\rm max}=2$, and $k_{\rm max}=1$).}
\label{fig_4}
\end{figure}

AFDM is an innovative multicarrier waveform that utilizes linear chirp signals as subcarriers. 
Its core principle involves mapping information symbols onto a twisted time-frequency domain (known as the DAFT domain), thereby achieving delay-Doppler orthogonality and enabling full diversity in doubly selective channels.

Fig. \ref{fig_3} illustrates the transceiver architecture of the AFDM-ICSC system.
At the transmitter, information bits are processed through channel coding, interleaving, and constellation mapping to generate data symbols.
The embedded pilot symbols for channel estimation are inserted into the data symbols and simultaneously serve as probing signals for channel sounding.
To prevent interference between pilot and data symbols, guard intervals are inserted between them, as shown in Fig. \ref{fig_4}.
The number of guard intervals, denoted by $Q$, is determined by the channel’s maximum normalized delay spread $l_{\rm max}$ and the maximum normalized Doppler shift $k_{\rm max}$.
AFDM multiplexes $ N $ information symbols onto $ N $ orthogonal chirp subcarriers via the inverse DAFT (IDAFT), which generates the discrete time-domain AFDM signal.
Due to the distinct periodicity of the signal generated by the IDAFT,
a chirp-periodic prefix (CPP) of length $L_{\rm cpp}$ is appended to the start of the discrete time domain signal. 
The CPP plays a similar role to the cyclic prefix (CP) used in OFDM, where the time-domain signal is generated via the inverse DFT (IDFT).
This design reveals a key structural distinction: the prefix type (CPP vs. CP) inherently follows from the underlying transform (IDAFT vs. IDFT), reflecting the signal periodicity of chirp- and tone-based waveforms.
Additionally, a preamble for time and frequency synchronization is added.
The resulting AFDM signal is then transmitted over the channel.

The received AFDM signal comprises $P$ replicas of the transmitted signal, each corresponding to a distinct path in the channel.
After synchronization and discarding the CPP, the received discrete time-domain AFDM signal is obtained. 
This signal is then demodulated using the DAFT, employing the same $c_1$ and $c_2$ parameters as in modulation, to recover the received symbols in the DAFT domain.
The received AFDM signal in the DAFT domain undergoes channel estimation and data detection, followed by constellation demapping, deinterleaving, and channel decoding to recover the transmitted bits. 
Consequently, communication metrics such as bit error rate (BER) and error vector magnitude (EVM) can be assessed. 
Furthermore, based on the statistical analysis of the channel estimation, channel characteristics can also be extracted including the PDP, DPS, RMS delay spread, and RMS Doppler spread.

Fig. \ref{fig_4a} illustrates how the embedded pilot facilitates channel estimation in AFDM.
For clarity, a single isolated pilot is inserted in the transmitter, with $Q$ guard intervals placed on both sides to mitigate interference caused by multipath and Doppler effects.
The light blue and light brown guard intervals are reserved to accommodate potential shifts in the pilot and data symbols, respectively.
Since the DAFT-domain impulse response provides a full representation of the channel in the delay-Doppler domain, AFDM localizes multipath components through threshold-based estimation across the $Q+1$ indices affected by the channel.
Thanks to its one-dimensional structure, AFDM requires nearly half the guard overhead of OTFS (see Fig. \ref{fig_4}), offering substantial benefits in multi-user scenarios.
Furthermore, AFDM directly extracts channel parameters—delay, Doppler, and path gain—without intermediate CIR or CFR estimation, making it inherently well suited for ICSC.
\begin{figure*}[!t]
\centering
\includegraphics[width=1\textwidth]{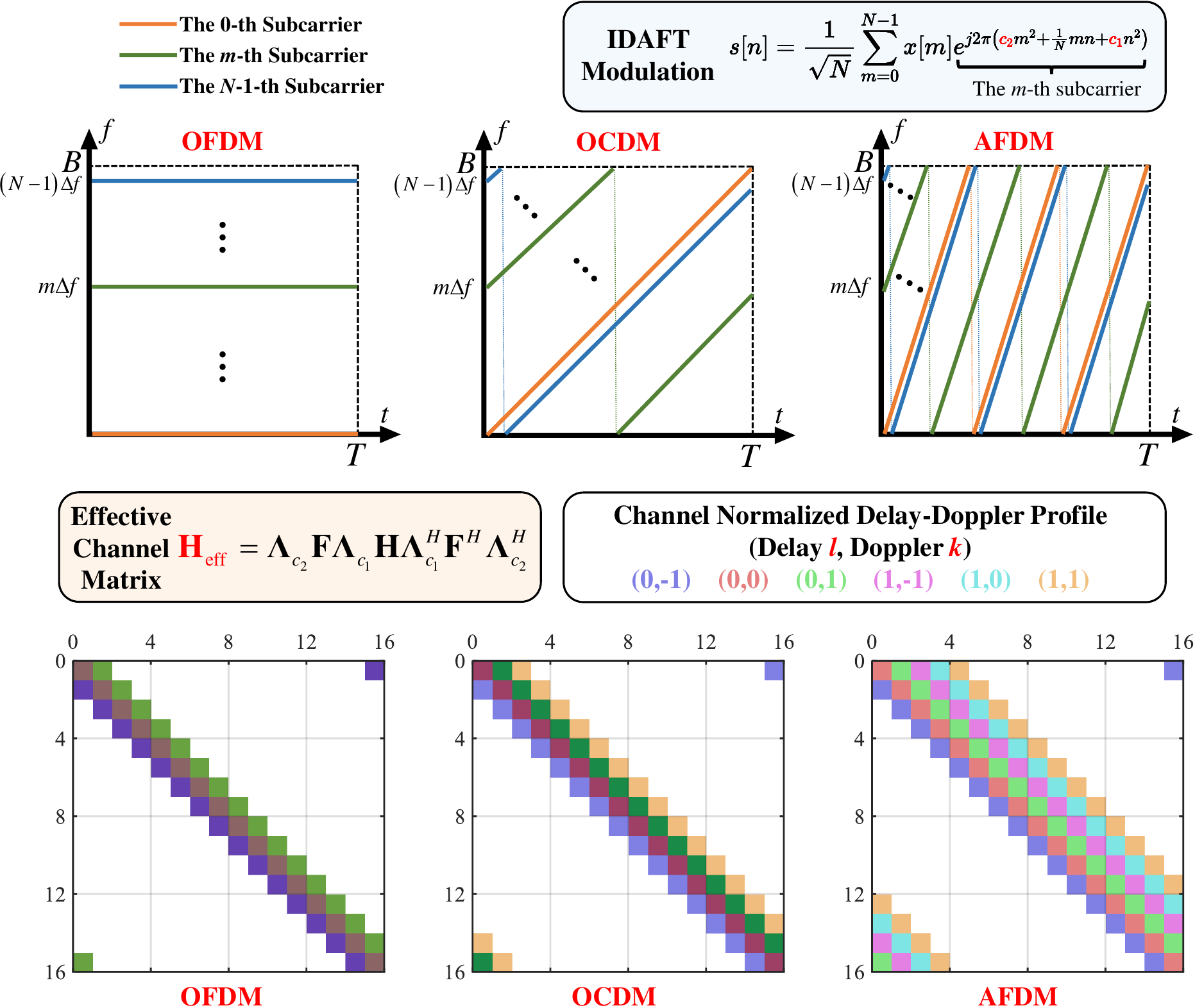}
\caption{Time-frequency representations and effective channel matrices of OFDM, OCDM, and AFDM ($N=16$, $P=6$, $l_{\rm max}=1$, and $k_{\rm max}=1$).}
\label{fig_5}
\end{figure*}

\subsection{Parameter Design Principles}
The IDAFT-based AFDM modulation formula is presented in the upper right corner of Fig. \ref{fig_5}, with the key parameters $c_1$ and $c_2$ highlighted in red.
Specifically, $c_1$ defines the slope of all chirp subcarriers, while $c_2 m^2$ determines the initial phase of the $m$-th subcarrier. 
The time-frequency representations of OFDM ($c_1=c_2=0$), OCDM ($c_1=c_2=\frac{1}{2N}$), and AFDM are presented in Fig. \ref{fig_5}, where $B$ represents the signal bandwidth, $\Delta f$ denotes the subcarrier spacing, and $T$ is the symbol duration.
This demonstrates that OFDM and OCDM are special cases of AFDM, emphasizing the compatibility of AFDM with these waveforms and its adaptability to different channel conditions through tunable parameters.
As demonstrated in \cite{ref10}, by selecting $c_1 \geq \frac{2(k_{\rm max}+\xi)+1}{2N}$ and setting $c_2$ to be an arbitrary irrational number or a rational number sufficiently smaller than $\frac{1}{2N}$, where $\xi$ represents the number of additional guard elements for fractional Doppler shifts, ensures that AFDM can achieve full diversity in doubly selective channels.
This capability is not achievable by fixed-parameter OFDM and OCDM.
To prevent interference between the pilot and data symbols, the number of guard intervals on each side of the pilot must satisfy $Q \geq 2Nc_1(l_{\rm max}+1)-1$.
A larger $c_1$ improves the resilience of AFDM to high mobility channels but increases the required guard interval to mitigate interference between pilot and data symbols.
Therefore, properly adjusting $c_1$ is crucial to balance communication performance and channel sounding overhead.
  
The matrix relationship between the input \( \textbf{x} \in \mathbb{C}^{N \times 1} \) and the output \( \textbf{y} \in \mathbb{C}^{N \times 1} \) in AFDM is given by $\textbf{y}=\textbf{H}_{\text{eff}} \textbf{x}+\textbf{w}$, where the equivalent channel matrix is defined as \( \textbf{H}_{\text{eff}} = \boldsymbol{\Lambda}_{c_{2}} \mathbf{F} \mathbf{\Lambda}_{c_{1}} \mathbf{H} \mathbf{\Lambda}_{c_{1}}^{H} \mathbf{F}^{H} \boldsymbol{\Lambda}_{c_{2}}^{H}\). 
Here, $\boldsymbol{\Lambda}_{c}=\operatorname{diag}\left(e^{-j 2 \pi c n^{2}}, n=0,1, \ldots, N-1\right)$ , $\textbf{H}$ denotes the time-domain channel matrix, $\textbf{F}$ is the DFT matrix, and $\textbf{w}$ is additive Gaussian noise.
The structure of \( \textbf{H}_{\text{eff}} \) is essential for accurate data detection at the receiver and serves as a basis for designing the AFDM parameters \( c_1 \) and \( c_2 \). 
By appropriately selecting $c_1$, $c_2$, and ensuring the channel satisfies the orthogonality condition $2(k_{\rm max}+\xi)(l_{\max}+1)+l_{\max}\leq N$, orthogonality in the normalized integer delay-Doppler domain of the AFDM equivalent channel can be guaranteed \cite{ref10}.
This orthogonality is crucial not only for ensuring optimal communication performance but also for mitigating interference from different paths during channel parameter estimation.
The position of the offset diagonals in the equivalent channel is determined by the normalized delay-Doppler integer indices of each path.
Each path generates a diagonal in the equivalent channel matrix, with its position determined by the normalized integer delay $l_p$ and Doppler shift $k_p$.
Specifically, in the equivalent channel matrices of OFDM, OCDM and AFDM, each path-induced diagonal is shifted by $k_p$, $l_p + k_p$, and $l_p\big(2k_{\max} + 1\big) + k_p$ positions, respectively.
A positive shift indicates a rightward displacement along the matrix diagonal, whereas a negative shift indicates a leftward displacement.
As shown in Fig. \ref{fig_5}, consider two paths characterized by delay-Doppler indices $(l_i,k_i)$ and $(l_j,k_j)$.
If $k_i=k_j$, the two paths overlap in the equivalent channel matrix of OFDM.
If $l_i+k_i=k_i+k_j$, the two paths overlap in the equivalent channel matrix of OCDM.
In contrast, AFDM allows all paths to be orthogonally distinguished by appropriately adjusting $c_1$.
Thus, by identifying these diagonal elements in the equivalent channel matrix, the delay and Doppler shifts of the channel can be precisely determined, providing an effective approach for channel parameter estimation.
\section{Potential applications and opportunities}
In this section, we explore the potential applications of the AFDM-ICSC, emphasizing its broad prospects and future development opportunities. 

\subsection{High-Frequency-Band Communication and Channel Sounding}
6G networks require support for higher data rates, necessitating the use of higher carrier frequencies (e.g., millimeter-wave, terahertz, and visible light bands) and wider system bandwidths.
However, channel characteristics vary significantly across these frequency bands.
In high-frequency bands, issues such as carrier frequency offset and phase noise are more pronounced, potentially disrupting the orthogonality of OFDM subcarriers and severely degrading communication performance.
Moreover, high-frequency wireless links experience severe path loss, while wider bandwidths exacerbate frequency-domain non-stationarity, intensifying frequency-selective fading.
To ensure stable communication and accurate channel estimation in high-frequency bands, waveforms with stronger coverage and improved robustness to carrier frequency offset are required (not offered by
OFDM).
In contrast, AFDM inherently provides superior Doppler resilience, along with enhanced robustness to carrier frequency offset and phase noise.
Moreover, the maximal time-frequency spreading of AFDM signals improves coverage, providing robustness against radio frequency impairments and enhancing adaptability and reliability in mitigating severe path loss.

\subsection{Vehicular Networks Communication and Channel Sounding}
Vehicular networks are a key application of 5G/6G in ultra-reliable low-latency communication scenarios, involving vehicle-to-vehicle, vehicle-to-infrastructure, and vehicle-to-pedestrian communications, collectively known as vehicle-to-everything (V2X).
In vehicular networks, the relative speeds between vehicles significantly reduce the coherence time of the channel, making it difficult for traditional channel sounding methods to accurately capture the characteristics of the channel varying over time.
As a result, the extracted channel features often fail to meet the requirements of real-time communication.
Additionally, the highly dynamic environment introduces severe Doppler spread, creating substantial challenges, particularly for OFDM, which is highly sensitive to frequency offsets and suffers performance degradation.
In contrast, AFDM provides a full delay-Doppler representation of the channel, enabling precise estimation of delay and Doppler parameters
while achieving full diversity gain.
Thus, AFDM-ICSC offers a feasible approach for vehicular networks, combining reliable communication with real-time channel sounding and laying a solid foundation for the future development of intelligent transportation systems and autonomous driving technologies.

\subsection{Non-Terrestrial Networks Communication and Channel Sounding}
6G networks will extend beyond traditional terrestrial communication to encompass non-terrestrial networks (NTN), such as satellite communication.
NTN provide global coverage, ubiquitous connectivity, and enhanced network reliability.
Satellite communication channels have unique characteristics, including large Doppler shifts, Doppler spread, frequency dependence, broad coverage, and long-range communication.
The high mobility of satellites causes time non-stationarity and large Doppler shifts, creating new challenges for channel sounding and communication.
AFDM has significant potential in NTN due to its strong ability to handle the Doppler effect.
Additionally, ICSC enables real-time channel sounding and continuous updating of channel models, effectively addressing issues arising from reduced channel coherence time.
Therefore, AFDM-ICSC provides a robust and efficient solution for reliable communication and channel sounding in NTN.

\subsection{Underwater Acoustic Communication and Channel Sounding}
Underwater acoustic (UWA) channels are among the most challenging wireless communication environments, which are characterized by non-Gaussian noise, limited frequency ranges, and severe path loss.
Refraction, reflection, and scattering further exacerbate severe multipath propagation.
UWA channels exhibit substantial dispersion in both time and frequency, resulting in time-variant and Doppler effects.
Traditional channel sounding techniques are limited by deployment challenges and high costs.
ICSC provides a promising solution to these challenges.
Although OFDM performs well in frequency-selective channels, its performance deteriorates significantly in UWA time-frequency selective channels due to inter-carrier interference.
In contrast, AFDM leverages strong delay-Doppler diversity, effectively adapting to the complex characteristics of UWA channels, and shows significant potential for both UWA communication and channel sounding.

\section{Future directions and challenges}
In this section, we outline the future research directions and key challenges of AFDM-ICSC systems.

\subsection{Synchronization}
Synchronization is crucial for maintaining the performance and reliability of ICSC systems, especially in complex communication environments.
At the medium access control
(MAC) layer, synchronization is vital for coordinating communications among multiple sensors and targets.
Without proper synchronization, devices may access the channel simultaneously, leading to signal interference, data collisions, and severe system performance degradation.
Consequently, designing efficient MAC protocols to improve device coordination under asynchronous conditions remains a major research challenge.
At the physical layer, synchronization is fundamental to reliable communication, especially in environments characterized by high mobility and rapidly varying channels.
Precise symbol synchronization is crucial for accurate data decoding.
However, the rapidly varying channel conditions in vehicular environments substantially complicate synchronization.
Additionally, dependence on known sequences for synchronization increases resource overhead.
Therefore, a key research focus is to develop synchronization techniques that offer high precision while minimizing resource overhead.

\subsection{Channel Estimation}
\begin{figure}[!t]
\centering
\subfloat[]{
\includegraphics[width=0.465\textwidth]{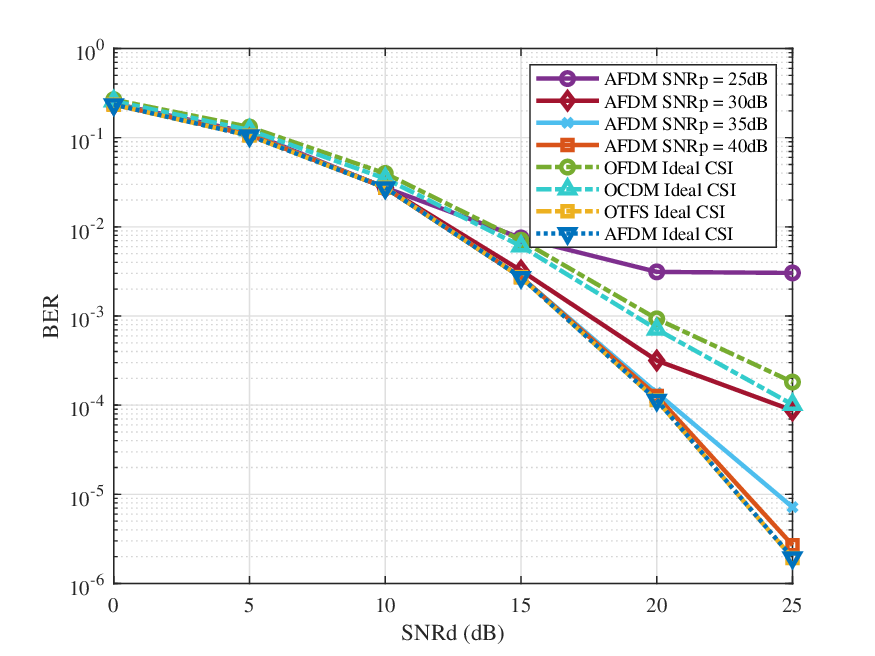}}\\
\subfloat[]{
\includegraphics[width=0.465\textwidth]{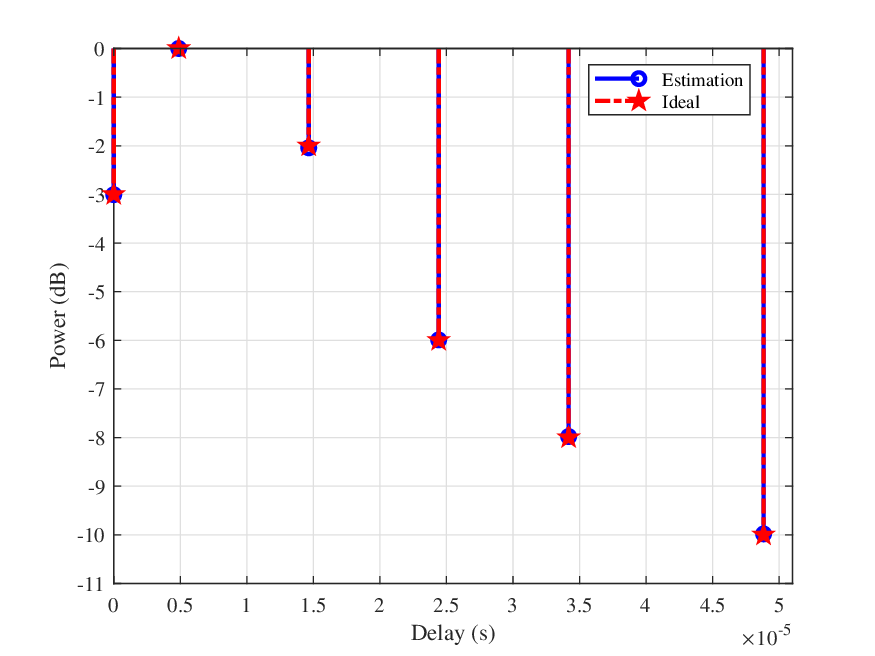}}\\
\subfloat[]{
\includegraphics[width=0.465\textwidth]{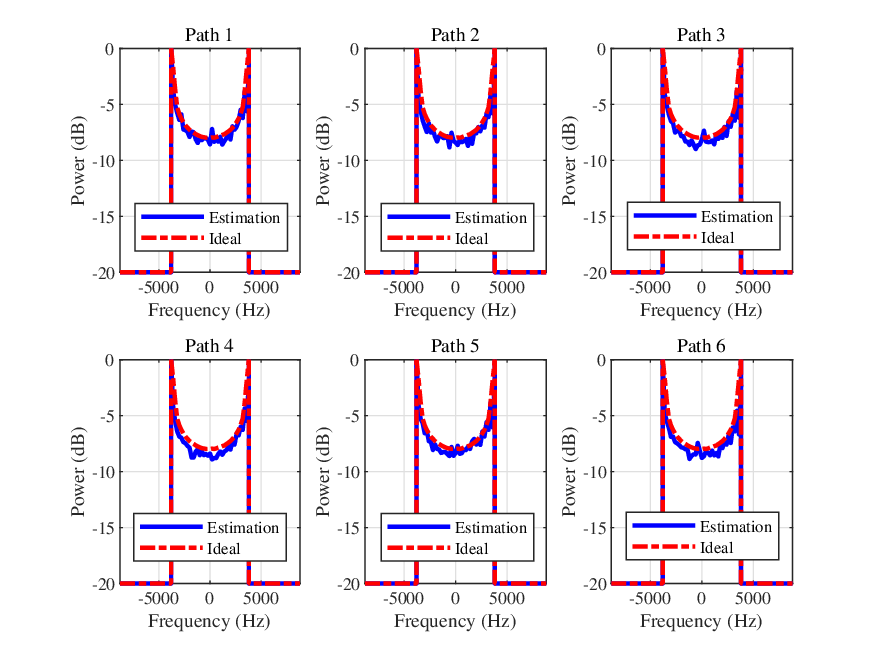}}
\caption{a) The BER performance versus data signal-to-noise ratio (SNRd) is compared for OFDM, OCDM, OTFS, and AFDM, with $N_{\rm OFDM} = N_{\rm OCDM} = N_{\rm AFDM} = 1024$, $M_{\rm OTFS} = N_{\rm OTFS} = 32$, using QPSK modulation and a linear minimum mean square error (LMMSE) detector; b) The PDP at SNRp = 35 dB and SNRd = 20 dB; c) The DPS at SNRp = 35 dB and SNRd = 20 dB, where the DPS of the six paths follows the Jakes distribution.}
\label{fig_6}
\end{figure}
Channel estimation is a fundamental challenge and a key research direction in practical communication systems, playing a crucial role in ICSC.
Unlike traditional time-domain CIR or frequency-domain CFR estimation, AFDM enables direct estimation of critical parameters, such as delay and Doppler shift, in the DAFT domain.
This is enabled by two key parameters, $c_1$ and $c_2$, which ensure delay-Doppler orthogonality and separability in the DAFT domain, facilitating precise channel representation.
This unique feature opens up new possibilities for AFDM-based channel sounding.
Research on channel estimation in AFDM-ICSC systems can be broadly categorized into two approaches. 
The first focuses on accurately extracting delay and Doppler parameters to characterize the channel. 
For example, the embedded pilot (EP) structure enables precise parameter estimation, as illustrated in Fig. \ref{fig_6}, where EP achieves near perfect BER and accurate PDP/DPS under high pilot signal-to-noise ratios (SNRp) \cite{ref10}. 
The second approach involves reconstructing the equivalent channel matrix to infer statistical properties without estimating individual parameters. A diagonal reconstruction method has been proposed to directly estimate the matrix, avoiding cumulative errors \cite{ref15}. 
In general, designing effective reference signals for accurate channel estimation remains a key challenge in ICSC, which requires a careful balance between communication and sounding resource allocation to optimize system performance.


\subsection{Privacy and Security}
As the Internet of Everything (IoE) evolves, privacy and security will become critical components of ICSC systems.
During communication and channel sounding, terminals frequently transmit reference signals for synchronization or sounding, which are vulnerable to interception by adversaries.
These signals can be exploited for unauthorized location tracking, leading to significant privacy risks to users.
Given these challenges, future research should prioritize the development of robust privacy protection mechanisms for these reference signals.
In this context, AFDM offers a unique advantage through its two modulation parameters, $c_1$ and $c_2$, enabling encrypted modulation designs that enhance the security of reference signals against unauthorized analysis \cite{ref12}.
Beyond this, the chirp-permuted AFDM variant further leverages these parameters to create a factorial-scaling search space, ensuring quantum-resilient communication even when an eavesdropper has perfect channel knowledge \cite{ref121}.
Additionally, developing effective attack detection and defense mechanisms is essential to mitigate common security threats, such as impersonation and man-in-the-middle attacks.
Collectively, these efforts will significantly enhance the security and robustness of ICSC systems, particularly in open, potentially adversarial environments.

\subsection{AI-Enabled Channel Modeling}
A key advantage of ICSC is its ability to leverage existing communication infrastructure to collect large-scale channel sounding data in diverse environments.
As wireless systems evolve toward higher frequency bands, larger antenna arrays, and more complex deployment environments, they face challenges such as exponential data growth and increasing channel complexity.
Traditional methods are increasingly limited in handling large-scale data and dynamic channel estimation.
Integrating artificial intelligence (AI) offers a new approach and a powerful solution.
With efficient data processing and robust learning capabilities, AI algorithms significantly improve performance in channel modeling and communication under complex conditions.
For example, clustering, classification, and regression algorithms can be used for multipath component clustering, scene classification, and channel prediction.
Furthermore, AI can extract higher-order features of dynamic channels, such as multipath clusters in complex environments.
Identifying these key features facilitates the design of intelligent algorithms and the optimization of communication systems, thereby improving overall system performance.

\section{Conclusions}
ICSC offers a cost-effective approach to real-time channel measurement and continuous updates, eliminating the need for dedicated sounding hardware.
Within this framework, AFDM distinguishes itself through robust delay–Doppler resilience and flexible adaptation enabled by two tunable chirp parameters, making it well suited for highly dynamic and doubly selective channels.
This article has provided a comprehensive overview of AFDM-ICSC, from its fundamental principles to application scenarios and key challenges.
Looking ahead, AFDM-ICSC is poised to become a cornerstone technology for 6G and beyond, delivering reliable communication and fine-grained channel characterization across diverse, high-mobility, and heterogeneous environments, thereby paving the way for intelligent, adaptive, and environment-aware wireless networks.


 


\end{document}